\documentclass[twocol]{ametsocV6_1}

\usepackage[hidelinks]{hyperref}
\usepackage{multirow}

\title{Evidence of mixed scaling for mean profile similarity \\ in the stable atmospheric surface layer}

\authors{Michael Heisel,\aff{a}\correspondingauthor{Michael Heisel,\\ michael.heisel@sydney.edu.au} 
Marcelo Chamecki\aff{b}
}

\affiliation{\aff{a}{School of Civil Engineering, University of Sydney, Sydney, NSW 2008, Australia}\\
\aff{b}{Department of Atmospheric \& Oceanic Sciences, University of California Los Angeles, Los Angeles, CA 90095, USA}
}

\abstract{A new mixed scaling parameter $Z=z/\sqrt{Lh}$ is proposed for similarity in the stable atmospheric surface layer, where $z$ is the height, $L$ is the Obukhov length, and $h$ is the boundary layer depth. Compared to the parameter $\zeta = z/L$ from Monin--Obukhov similarity theory (MOST), the new parameter $Z$ leads to improved mean profile similarity for wind speed and air temperature in large-eddy simulations. It also yields the same linear similarity relation for CASES-99 field measurements, including in the strongly stable (but still turbulent) regime where large deviations from MOST are observed. Results further suggest that similarity for turbulent energy dissipation rate depends on both $Z$ and $\zeta$. The proposed mixed scaling of $Z$ and relevance of $h$ can be explained by physical arguments related to the limit of z-less stratification that is reached asymptotically above the surface layer. While the presented evidence and fitted similarity relations are promising, the results and arguments are limited to a small sample of idealized stationary stable boundary layers. Corroboration is needed from independent datasets and analyses, including for complex and transient conditions not tested here.} 

\begin{document}

\maketitle

\section{Introduction}

The wind speed and air temperature above the Earth's surface are critical to both human activities and the interaction between Earth and the atmosphere. Accordingly, similarity in the shape of the mean profiles for these quantities is historically one of the most researched topics in the study of the atmospheric boundary layer (ABL). Under neutral conditions in the absence of a temperature gradient and other buoyancy effects, the wind speed is well described by the logarithmic law of the wall -- or simply the ``log law'' -- that can be derived using a variety of methods \citep[see, e.g.,][]{Karman1930,Prandtl1932,Millikan1938,Townsend1976}. A recent experimental assessment is provided by \citet{Marusic2013}. In its gradient form, the log law is

\begin{equation}
\frac{\partial \overline{U}}{\partial z} = \frac{u_*}{\kappa z},
\label{eq:loglaw}
\end{equation}

\noindent where $\overline{U}$ is the mean horizontal wind speed, $u_*$ is the friction (shear) velocity, $\kappa$ is the von K\'{a}rm\'{a}n constant, and $z$ is height above the surface. The log law is strictly valid in the lowest portion of the ABL known as the surface layer or inertial layer. While the surface layer is the focus of the present work, alternative incomplete similarity relations such as a wind speed power law are also common in various applications such as pollutant transport \citep{Arya1998}, wind energy \citep{Manwell2009}, and wind engineering \citep{Simiu2019} due to improved accuracy at higher positions above the surface layer \citep{Barenblatt1993}.

The prevailing framework to account for effects of temperature and buoyancy in the surface layer profiles comes from Monin--Obukhov similarity theory \citep[MOST,][]{Monin1954}. The theory postulates that the dimensionless gradients for wind speed (momentum)

\begin{equation}
\phi_m = \frac{\partial \overline{U}}{ \partial z} \, \left( \frac{\kappa z}{ u_*} \right) = f_m(\zeta)
\label{eq:mom}
\end{equation}

\noindent and air temperature (heat)

\begin{equation}
\phi_h = \frac{\partial \overline{\theta}}{ \partial z} \, \left( \frac{\kappa z}{ \theta_*} \right) = f_h(\zeta)
\label{eq:heat}
\end{equation}

\noindent are universal functions of the similarity parameter $\zeta = z/L$. The log law scaling in Eq. \eqref{eq:loglaw} is used to normalize the gradients such that $\phi_m = 1$ in the absence of buoyancy. The Obukhov length \citep{Obukhov1946}

\begin{equation}
L = \frac{u_*^2 \overline{\theta}}{\kappa g \theta_*}
\label{eq:obukhov}
\end{equation}

\noindent represents the height at which buoyancy mechanisms become important relative to shear based on scaling arguments for the turbulent kinetic energy (TKE) budget. Here, $g$ is the gravitational constant. The functions $f_m(\zeta)$ and $f_h(\zeta)$ quantify deviations of the mean profiles from the log law scaling and also connect the gradients to the surface fluxes for momentum $\overline{u^\prime w^\prime}_s = -u_*^2$ and heat $\overline{w^\prime \theta^\prime}_s = -u_* \theta_*$, respectively. MOST assumes that $z$ and $L$ are the only relevant length scales for the gradients in Eqs. \eqref{eq:mom} and \eqref{eq:heat}, such that additional length scales including the surface roughness and ABL depth have negligible impact on the surface layer gradients.

\begin{figure*}
\centerline{\includegraphics[width=39pc]{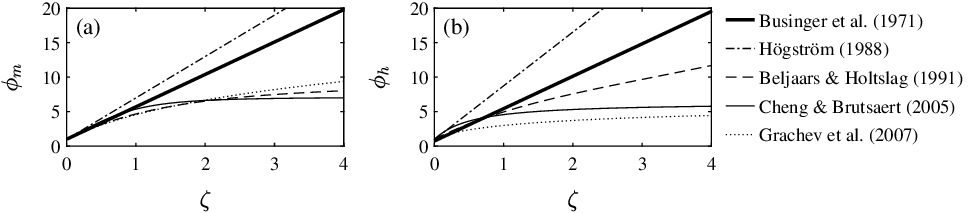}}
  \caption{Comparison of empirical similarity relations proposed in the literature for the stably-stratified surface layer. The relations are used to predict the dimensionless mean gradient for momentum $\phi_m = \partial \overline{U} / \partial z \, (\kappa z / u_*)$ (a) and heat $\phi_h = \partial \overline{\theta} / \partial z \, (\kappa z / \theta_*)$ (b). Each relation is a function of the Monin--Obukhov similarity parameter $\zeta = z/L$, where $L$ is the Obukhov length.}
  \label{fig1}
\end{figure*}

For the stable atmospheric boundary layer (SBL) in weak stratification ($0 < \zeta \lesssim 1$), the consensus from early field experiments is that $f_m(\zeta)$ and $f_h(\zeta)$ are linear functions \citep{McVehil1964,Zilitinkevich1968,Webb1970,Oke1970,Businger1971,Dyer1974} that result from a power series expansion in the original theory \citep{Monin1954}. Upon integration of the gradient relations, the resulting mean profiles comprise a combination of logarithmic and linear components, where the linear contribution becomes dominant for large $\zeta$ in the limit of z-less stratification \citep{Wyngaard1972}. However, numerous experiments have shown $\phi$ to deviate from the linear relations under increased stratification and become a weaker function of $\zeta$ for $\zeta > 1$ \citep[][among others]{Webb1970,Beljaars1991,Howell1999,Grachev2005,Ha2007,Optis2014}. These observations have led to more complex functional forms for $f_m(\zeta)$ and $f_h(\zeta)$ that are approximately linear for weak stratification and capture the flattened trends for large $\zeta$. A comparison of several empirical relations fitted to a variety of experimental datasets is shown in Fig. \ref{fig1} \citep{Businger1971,Hogstrom1988,Beljaars1991,Cheng2005,Grachev2007}.

Observed differences across the empirical relations in Fig. \ref{fig1} are often attributed to experimental uncertainty \citep{Yaglom1977}, the passage of large-scale eddies \citep{Salesky2020}, or the use of data periods in conditions where MOST is not valid \citep{Kouznetsov2010}. The latter point is particularly relevant for very stable conditions characterized by weak winds and global intermittency \citep{Holtslag1986,Mahrt1999}. Previous works have shown that data points deviating from the linear trends are often associated with conditions exhibiting nonstationarity \citep{Mahrt2007} or a non-canonical energy spectrum without a clear inertial subrange \citep{Grachev2013}.

It is also possible that some of the variability in empirical fits is due to the relevance of additional parameter(s) such that $\zeta$ provides incomplete similarity. Zilitinkevich and co-authors used asymptotic matching arguments and generalized length scales to introduce dependencies on the Coriolis frequency $f$ and properties of the capping inversion such as the Brunt-V\"{a}is\"{a}l\"{a} frequency $N$ \citep{Zilitinkevich1989,Zilitinkevich2000,Zilitinkevich2005}. The length scales associated with these frequencies, i.e. $u_*/f$ and $u_*/N$ \citep{Zilitinkevich2005}, are closely related to the equilibrium ABL depth $h$ \citep{Kitaigorodskii1988}, which suggests $\phi$ may depend on $h$ in addition to $L$. Previous studies have introduced $h$ into the similarity relations, but have predominately focused on augmenting MOST to capture trends above the surface layer \citep[e.g.,][]{Gryning2007,Optis2014}.

There is further research into the influence of $h$ on stably-stratified turbulence based on the transition between continuous and intermittent turbulence. Multiple parameters and knowledge of the overall boundary layer structure are required to predict the transition between these regimes \citep[see, e.g.,][]{Williams2013,Monahan2015}. In particular, simulations have shown that the critical point for the collapse of turbulence depends on both the stability and a bulk Reynolds number \citep{Deusebio2015}, and in certain cases $h/L$ is explicitly used \citep{Wiel2007,Donda2014}. The role of $h$ in the collapse of turbulence suggests it is a critical parameter to the phenomenology of stratified flows, and this idea is extended here to the fully turbulent regime through a reassessment of structural similarity.

In consideration of the variability seen in Fig. \ref{fig1} and previous research into the importance of the boundary layer depth, the goal of the present study is to account for surface layer trends by expanding the parameter space of the similarity functions $\phi = f(z,L,h)$ to include $h$. Specifically, a revised similarity parameter $Z = z / \sqrt{L h}$ is proposed based on a composite length scale using the geometric mean of $L$ and $h$. It will be shown that $Z$ provides improved similarity compared to traditional MOST for high-resolution large-eddy simulations (LES) \citep{Sullivan2016} and yields a linear trend extending to strongly stable (but still turbulent) conditions in measurements from the Cooperative Atmosphere-Surface Exchange Study--1999 (CASES-99) field campaign \citep{Poulos2002}. A physical basis for the composite length scale $\sqrt{Lh}$ is proposed that considers the profiles to asymptotically approach a z-less gradient above the surface layer and utilizes observed stability relations for resistance laws.

A brief overview of the LES and CASES-99 measurements is provided in section \ref{sec:data}. A comparison of $\phi ( \zeta )$ and $\phi ( Z )$ is then presented in section \ref{sec:results}. A justification for the mixed scaling and the connection between $\phi(\zeta)$ and $\phi(Z)$ are discussed in section \ref{sec:discuss}. Concluding remarks are given in section \ref{sec:conclude}. A detailed account of the CASES-99 data analysis is reserved for the appendix.

\section{Stable atmospheric boundary layer data}
\label{sec:data}

The following descriptions of the LES and field measurements in the subsections below are limited to a summary of the most relevant details. The LES was previously presented elsewhere \citep{Sullivan2016,Heisel2023}, and the CASES-99 field campaign has been discussed in numerous studies \citep[see, e.g.,][among many others]{Poulos2002,Sun2002,Banta2002,Cheng2005,Ha2007}. A full account of details required to reproduce the analysis of CASES-99 measurements is given in the appendix.

\subsection*{Large-eddy simulations}
\label{subsec:les}

The LES domain design and imposed boundary conditions are based on the GEWEX Atmospheric Boundary Layer Study (GABLS) benchmark case \citep[see, e.g.,][]{Beare2006}.  Four stability cases were achieved by applying a fixed surface cooling rate that ranged from $C_r=$ 0.25 to 1 K\,h$^{-1}$ across the different cases. The results presented here are based on average flow statistics between 8 and 9 physical hours of simulation when the ABL has reached near-equilibrium conditions except for a constant temperature shift owing to the fixed surface cooling \citep{Sullivan2016}. The conditions represent a simplified canonical case of a long-lived stable ABL. The 1024$^3$ numerical grid for each case corresponds to an isotropic resolution of $\Delta = $ 0.39 m, noting the effective horizontal resolution is coarser due to the dealiasing scheme \citep{Sullivan2016}. The stable ABL cases are supplemented by a conventionally neutral simulation on a 512$^3$ grid \citep{Heisel2023}. The neutral case imposed zero surface heat flux and a stable capping inversion, such that weak buoyancy effects are present in the surface layer due to entrainment and downward propagation from the inversion. The conventionally neutral case is only included for later results that use local-in-height scaling.

The flux statistics for momentum $\overline{ u^\prime w^\prime}$ and kinematic temperature $\overline{ w^\prime \theta^\prime}$ are based on the sum of the resolved and subgrid-scale components. Further, the velocity statistics $\overline{U}$ and $\overline{ u^\prime w^\prime}$ are calculated as the magnitude of the horizontal components along $x$ and $y$ to account for the moderate wind veer that is present in the surface layer. The ABL depth $h = z({-}\overline{u^\prime w^\prime}{=}0.05u_*^2)/0.95$ is estimated based the height where the average shear stress is 5\% of the surface value \citep{Kosovic2000}. The relevant scaling parameters for the conventionally neutral and stably-stratified ABL are given in Table \ref{tbl1}. In later figures, the cases are classified based on the stability parameter $h/L$.

\begin{table}
\caption{Key scaling parameters from large-eddy simulations (LES) of the atmospheric boundary layer (ABL) under conventionally neutral \citep{Heisel2023} and stably-stratified \citep{Sullivan2016} conditions. Here, $h$ is the ABL depth based on the flux profile decay, $L$ is the Obukhov length, $u_*$ is the friction velocity, $\theta_*$ is the surface temperature scaling, and $\Delta_z$ is the LES grid resolution along the vertical direction.}\label{tbl1}
\begin{center}
\begin{tabular}{cccccc}
\topline
$h/L$	& $h$	& $L$		& $u_*$ 			& $\theta_*$	& $\Delta_z/h$	\\
(--)		& (m)	& (m)		& (m\,s$^{-1}$)	& (K)		& (--)			\\
\midline
0.0 		& 258 	& $O(10^5)$	& 0.33			& 0.000		& 0.0030			\\
1.4		& 160	& 116 		& 0.26 			& 0.038 		& 0.0024			\\
1.8		& 136 	& 75			& 0.23			& 0.049		& 0.0029			\\
2.2 		& 122	& 55			& 0.22			& 0.061		& 0.0032			\\
3.5		& 89		& 26			& 0.19			& 0.100		& 0.0044			\\
\botline
\end{tabular}
\end{center}
\end{table}

Gradients of the mean profiles are computed using centered finite difference evaluated at the midheights between vertical grid points. Later results exclude statistics from the lowest 3\% of the ABL relative to $h$, which is approximately the lowest 10 vertical grid points. Statistics within the excluded region exhibit so-called ``overshoot'' and a clear dependence on the wall model \citep{Mason1992,Brasseur2010}. The number of excluded grid points will depend on the employed subgrid-scale and wall models, such that the appropriate cutoff should be evaluated independently in future studies.

\subsection*{CASES-99 field measurements}
\label{subsec:cases}

The CASES-99 field campaign occurred during October 1999 over flat grassland in southeastern Kansas \citep{Poulos2002}. Numerous measurement systems were deployed, including a 60-meter-tall meteorological (met) tower. The tower was instrumented with eight sonic anemometers ranging from 1.5 m to 55 m above ground level, six vane anemometers, six thermistors, and thirty-four thermocouples.

Mean and turbulent statistics are calculated from 5-minute data periods that meet quality control criteria. Turbulent flux statistics are computed exclusively using the sonic anemometers due to the high sampling rate of 20 Hz and their simultaneous measurement of velocity and temperature. The scaling parameters $u_*$, $\theta_*$, and $h$ are estimated by fitting idealized nondimensional flux profiles \citep{Nieuwstadt1984} to the measurements. The fitting procedure limits the analysis to conditions with a decaying flux across the height of the met tower, thus excluding periods exhibiting pronounced ``top-down'' turbulence and other more complex vertical structure \citep{Mahrt2022}. The appendix material outlines in further detail the quality control criteria, flux profile fitting procedure, and methods employed to estimate flux-gradient statistics.

\section{Results}
\label{sec:results}

\subsection*{Surface scaling}

\begin{figure*}
\centerline{\includegraphics[width=39pc]{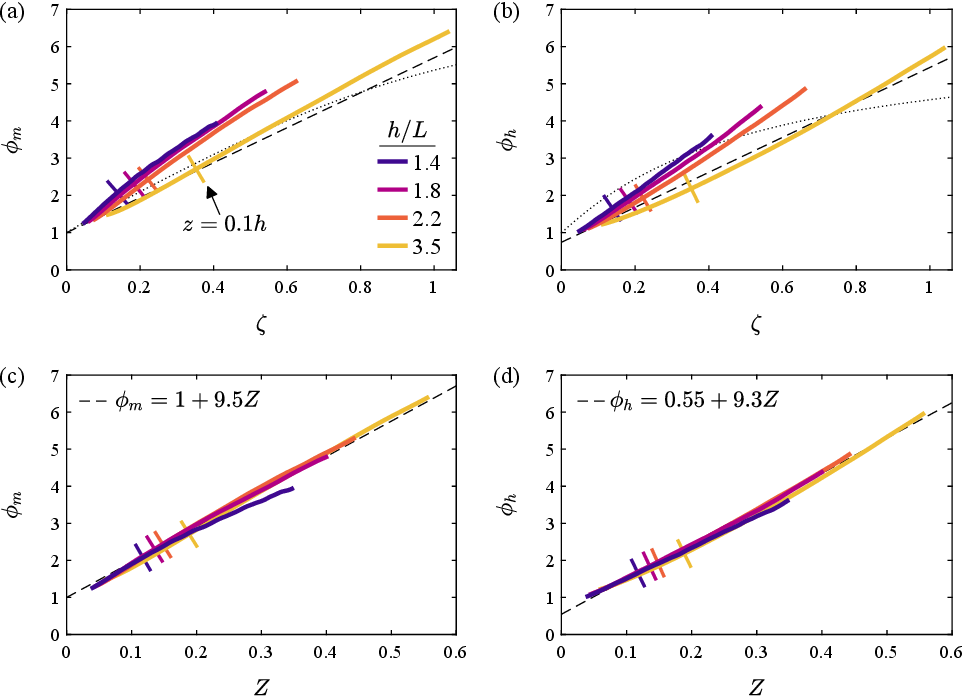}}
  \caption{Comparison of the traditional similarity parameter $\zeta=z/L$ and the proposed parameter $Z=z/\sqrt{Lh}$ for the simulations of the stable ABL in Table \ref{tbl1}: (a) $\phi_m(\zeta)$; (b) $\phi_h(\zeta)$; (c) $\phi_m(Z)$; and (d) $\phi_h(Z)$. Vertical positions from $z=$ 0.03$h$ to 0.3$h$ are included here, and the short lines indicate the height $z=0.1h$ for reference. The reference lines in (a,b) are the $\phi(\zeta)$ expressions from \citet[][dashed]{Businger1971} in Eq. \eqref{eq:businger} and \citet[][dotted]{Cheng2005} in Eq. \eqref{eq:cheng}. The relations in (c,d) result from a least-squares fit to the LES. Different colors in this and later figures correspond to variability in $h/L$.}
  \label{fig2}
\end{figure*}

The nondimensional gradients for momentum in Eq. \eqref{eq:mom} and heat in Eq. \eqref{eq:heat} are shown as a function of $\zeta$ in Figs. \ref{fig2}(a) and \ref{fig2}(b), respectively, for the LES cases in Table \ref{tbl1}. The conventionally neutral case is excluded here for the evaluation of surface scaling, but is included in later results. Surface scaling refers to the traditional definition of $\phi$ and $L$ using surface fluxes $u_*$ and $\theta_*$. The value $\kappa = 0.4$ is used for the von K\'{a}rm\'{a}n constant. Heights between $z=$ 0.03$h$ and 0.3$h$ are included in Fig. \ref{fig2} and later plots. The maximum height is above the traditional limit of the surface layer (i.e. 0.1$h$, indicated for reference in each figure). The range is extended to 0.3$h$ because the observed surface layer trends continue to this height, including a local equilibrium in the TKE budget seen in later results.

Two empirical similarity relations from Fig. \ref{fig1} are included in Fig. \ref{fig2}(a,b) for reference.  The relations from \citet[][dashed lines]{Businger1971} are

\begin{equation}
\begin{split}
\phi_m(\zeta) &= 1+4.7 \zeta \\
\phi_h(\zeta) &= 0.74+4.7 \zeta.
\end{split}
\label{eq:businger}
\end{equation}

\noindent The function from \citet[][dotted lines]{Cheng2005} is

\begin{equation}
\phi(\zeta) = 1 + a \left( \frac{ \zeta + \zeta^b \left( 1 + \zeta^b \right) ^{\tfrac{1-b}{b}} }{ \zeta + \left( 1 + \zeta^b \right) ^{\tfrac{1}{b}} } \right),
\label{eq:cheng}
\end{equation}

\noindent where ($a=$ 6.1, $b=$ 2.5) for $\phi_m$ and ($a=$ 5.3, $b=$ 1.1) for $\phi_h$. The expression is designed to account for observed trends in the strongly stable regime ($\zeta > 1$) and is used in atmospheric models such as the Weather Research and Forecasting model \citep{Jimenez2012}.

The MOST relations capture a majority of the deviation in the profiles from the neutral log law scaling. See, for example, Fig. 1 of \citet{Heisel2023}. However, the residual differences in Fig. \ref{fig2}(a,b) exhibit a clear stability trend. The order of the cases, i.e. decreasing $\phi$ with increasing bulk stability, suggests that MOST overcompensates for the effect of stable stratification on the gradients for the conditions simulated by the LES. Previous works have identified similar stability trends and noted the possibility for an $h/L$ correction in convective conditions \citep{Khanna1997,Johansson2001,Salesky2012}.

The similarity can be improved by weakening the dependence on $L$ in the revised parameter $Z=z/\sqrt{Lh}$. Incorporating the SBL depth $h$ in $Z$ is another critical component for the collapse of profiles throughout the surface layer and up to 0.3$h$ as seen in Fig. \ref{fig2}(c,d). For instance, using $h$ defined from the flux profiles leads to greater similarity compared to alternate representations of the SBL depth such as the inversion layer height $z_i$ (not shown) that varies more weakly with stratification. The profiles exhibit weak curvature, but are well approximated by the fitted linear relations

\begin{equation}
\begin{split}
\phi_m(Z) &= 1+9.5 Z \\
\phi_h(Z) &= 0.55+9.3 Z.
\end{split}
\label{eq:heist}
\end{equation}

\noindent The intercept of 1 for $\phi_m(Z)$ is imposed to match the log law for neutral conditions, and is 0.9 if the fit is unconstrained. The intercept for $\phi_h$ is 0.72 if only heights below 0.1$h$ are considered, where the difference is due to the weak convex curvature across the extended range of heights. The interpretation of the fitted intercept and slope values is further discussed later. The more important outcome of Fig. \ref{fig2} is the markedly improved collapse of profiles corresponding to $\phi(Z)$.

Table \ref{tbl2} quantifies the agreement between the observed $\phi$ values and those predicted by the empirical relations featured in Fig. \ref{fig2}. The normalized mean squared error \citep[NMSE,][]{Chang2004} measures the goodness of fit, and the correlation coefficient $\rho$ is a metric for data scatter irrespective of systematic error. The low NMSE and $\rho>$ 0.99 values for the revised similarity parameter are consistent with the close overlap of profiles seen in Fig. \ref{fig2}(c,d).

\begin{table*}
\caption{The normalized mean squared error (NSME, right columns) and correlation coefficient ($\rho$, right columns) quantifying the alignment of the observed $\phi$ values with those predicted by empirical similarity relations: surface scaling for $\phi_m$ and $\phi_h$ in Figs. \ref{fig2} and \ref{fig3}; local scaling $\Phi_m$ and $\Phi_h$ in Fig. \ref{fig4}; and local scaling for dissipation $\Phi_\epsilon$ in Fig. \ref{fig5}. The similarity relations are defined in Eq. \eqref{eq:businger} for \citet{Businger1971}, in Eq. \eqref{eq:cheng} for \citet{Cheng2005}, and in the corresponding figures for the proposed parameter $Z=z/\sqrt{Lh}$.
}\label{tbl2}
\begin{center}
\begin{tabular}{rrccc|ccc}
\topline
	&	& \multicolumn{3}{c}{\textbf{NMSE}}	& \multicolumn{3}{c}{\textbf{correlation}}	\\
	&	& Businger	& Cheng and	& $f(Z)$	& Businger	& Cheng and	& $f(Z)$	\\
	&	& et al. \citeyearpar{Businger1971}	& Brutsaert \citeyearpar{Cheng2005}	& (present)	& et al. \citeyearpar{Businger1971}	& Brutsaert \citeyearpar{Cheng2005}	& (present)	\\
\midline
\multirow{5}{*}{\rotatebox[origin=c]{90}{\parbox[c]{1cm}{\centering LES}}}	& $\phi_m$	& 0.059	& 0.035	& 0.001	& 0.96	& 0.97	& $>$0.99	\\
	& $\phi_h$	& 0.041	& 0.029	& 0.001	& 0.96	& 0.96	& $>$0.99	\\
	& $\Phi_m$	& 0.059	& 0.054	& 0.001	& 0.97	& 0.98	& $>$0.99	\\
	& $\Phi_h$	& 0.031	& 0.035	& 0.001	& 0.97	& 0.95	& $>$0.99	\\
	& $\Phi_\epsilon = \Phi_m-\zeta$	& 0.072	& 0.066	& 0.003	& 0.95	& 0.96	& $>$0.99	\\
\midline
\multirow{4}{*}{\rotatebox[origin=c]{90}{\parbox[c]{1.5cm}{\centering CASES-99}}}	& $\phi_m$	& 0.46	& 0.23	& 0.13	& 0.89	& 0.87	& 0.93	\\
	& $\phi_h$	& 0.73	& 0.35	& 0.25	& 0.77	& 0.81	& 0.84	\\
	& $\Phi_m$	& 0.95	& 0.35	& 0.12	& 0.87	& 0.83	& 0.93	\\
	& $\Phi_h$	& 1.98	& 0.38	& 0.29	& 0.73	& 0.77	& 0.81	\\
\botline
\end{tabular}
\end{center}
\end{table*}

\begin{figure*}
\centerline{\includegraphics[width=39pc]{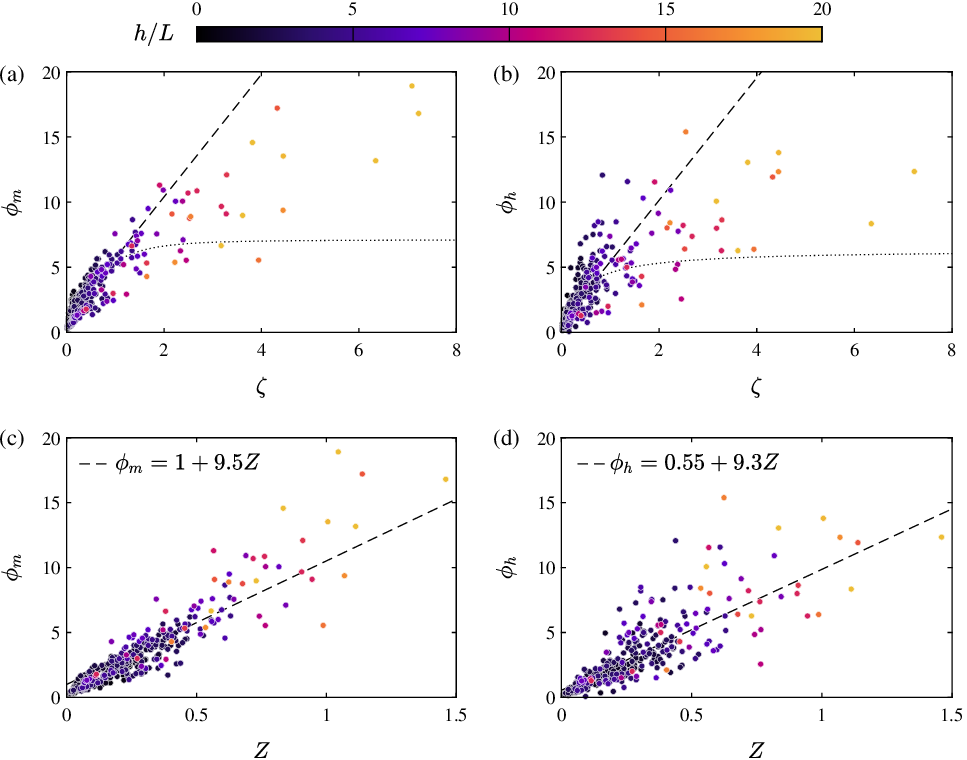}}
  \caption{Comparison of the traditional similarity parameter $\zeta=z/L$ and the proposed parameter $Z=z/\sqrt{Lh}$ for CASES-99 field measurements in stable conditions: (a) $\phi_m(\zeta)$; (b) $\phi_h(\zeta)$; (c) $\phi_m(Z)$; and (d) $\phi_h(Z)$. Vertical positions up to $z=$ 0.3$h$ are included here, and marker color indicates the estimated $h/L$ value. The reference lines in (a,b) are the $\phi(\zeta)$ expressions from \citet[][dashed]{Businger1971} in Eq. \eqref{eq:businger} and \citet[][dotted]{Cheng2005} in Eq. \eqref{eq:cheng}. The relations in (c,d) correspond to linear fits to the LES shown in Fig. \ref{fig2}(c,d) and Eq. \eqref{eq:heist}.}
  \label{fig3}
\end{figure*}

Figure \ref{fig3} shows the same statistics as Fig. \ref{fig2}, except the CASES-99 field measurements are featured rather than the LES. Figure \ref{fig3} employs the same upper height limit of $z=0.3h$ as before. A significantly wider range of $\zeta$ and $Z$ values is observed relative to Fig. \ref{fig2}, thus extending the similarity scaling comparison from weak and moderate stratification to more stable conditions. The wider range is also apparent from the estimated bulk stability $h/L$ indicated by the color of each data marker in Fig. \ref{fig3}, where the range of $h/L$ values is due to large variability in both $h$ and $L$ across 5-minute periods (see, e.g., Fig. \ref{figA2}). Traditional MOST and the proposed mixed scaling parameter are related as $Z = \zeta (h/L)^{-1/2}$, such that the difference between $\zeta$ and $Z$ increases with increasing $h/L$. This trend is observed in Fig. \ref{fig3}(a,b), where the deviation from a linear regression is predominately due to strongly stable periods with large $h/L$. Further, the flattening of the trend with increasing $h/L$ yields a curve resembling the non-linear functions in Fig. \ref{fig1}. The curvature results in high NMSE values in Table \ref{tbl2} for the linear \citet{Businger1971} relations.

The mixed scaling parameter $Z$ in Fig. \ref{fig3}(c,d) compensates for the observed $h/L$ dependence and produces a $\phi(Z)$ relation that is approximately linear within the scatter of the data markers. Based on the consistently lower NMSE and higher $\rho$ values in Table \ref{tbl2}, the expressions in Eq. \eqref{eq:heist} align well with the field measurements, despite being fitted to the LES data that is confined to lower $\zeta$ and $Z$ values. The alignment of Eq. \eqref{eq:heist} with the LES and CASES-99 data across a wide range of stable stratification and the reduction in scatter suggest that the composite length $\sqrt{Lh}$ is a relevant scaling parameter for the mean profiles in both weakly- and strongly-stratified conditions, excluding the regime of intermittent turbulence that is not assessed here.

The similarity of $\phi(Z)$ for weak stratification is specific to the LES results in Fig. \ref{fig2}(c,d). For the field data in Fig. \ref{fig3}, the scatter and uncertainty of the results exceed any differences between $\zeta$ and $Z$ for weakly stable conditions (e.g. $\zeta<$ 1). Further, many of the points in this regime appear to fall below the reference curves. The analysis outlined in the appendix is not optimized for identifying and assessing periods of weak stratification, particularly when $h$ is well above the tower height. Longer data periods, stricter stationarity criteria, and an alternative procedure for estimating $h$ would all help to refine the statistics for weak stratification. In this context, conclusions drawn from Fig. \ref{fig3} are limited to the pronounced and well-defined trends for large $h/L$.

\begin{figure*}
\centerline{\includegraphics[width=39pc]{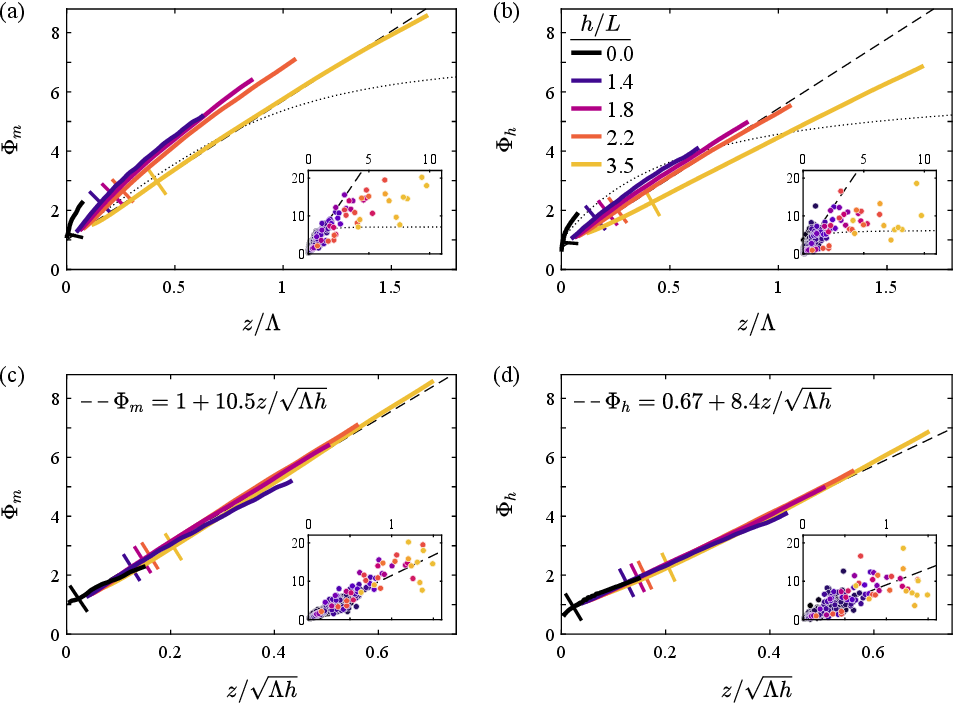}}
  \caption{Same as Figs. \ref{fig2} and \ref{fig3}, except the dimensionless gradients $\Phi$ and Obukhov length $\Lambda$ are defined using local-in-height fluxes rather than surface parameters, including for the conventionally neutral LES case. Vertical positions from $z=$ 0.03$h$ to 0.3$h$ are included here, and the short lines indicate the height $z=0.1h$ for reference. The reference lines in (a,b) are the $\phi(\zeta)$ expressions from \citet[][dashed]{Businger1971} in Eq. \eqref{eq:businger} and \citet[][dotted]{Cheng2005} in Eq. \eqref{eq:cheng}. The relations in (c,d) result from a least-squares fit to the LES.}
  \label{fig4}
\end{figure*}

One caveat to the trends in Fig. \ref{fig3} is self-correlation due to shared terms in the nondimensional parameters \citep{Kenney1982}, e.g. $u_*$ appears in both $\phi_m$ and $\zeta$.  To test this effect, the values for the dimensional gradients $\partial \overline{U}/\partial z$ and $\partial \overline{\theta}/\partial z$ were randomized across the observations \citep{Hicks1981,Klipp2004}, while fixing the scaling parameters to preserve their correlations. The randomized momentum results yielded similar levels of self-correlation: $\rho \approx$ 0.7 for $\zeta$ and 0.75 for $Z$. However, the randomized heat values were weakly correlated for all cases, i.e. $\rho < 0.15$. While self-correlation likely contributes to part of the reduced scatter for momentum, it explains neither the high correlations for $\phi_h$ nor the alignment of Eq. \eqref{eq:heist} with both LES and CASES-99 data.

\subsection*{Local scaling}

Many of the data points in Fig. \ref{fig3}, particularly for large $h/L$, correspond to the stability regime and range of heights where local-in-height scaling is recommended \citep{Nieuwstadt1984,Sorbjan1986}. Accordingly, the results in Figs. \ref{fig2} and \ref{fig3} are re-evaluated in Fig. \ref{fig4} using local scaling parameters, including for the conventionally neutral LES case. Specifically, the nondimensional gradients $\Phi$ and local Obukhov length $\Lambda(z)$ are now defined using $\overline{u^\prime w^\prime}(z)$ and $\overline{w^\prime \theta^\prime}(z)$ in place of $u_*$ and $\theta_*$. For the CASES-99 results in the inset panels, the local fluxes correspond to those computed directly from the sonic anemometers, not the value of the fitted flux profile at the same height.

As seen in Fig. \ref{fig4}(a,b), local scaling does not account for the $h/L$ trends observed for $\phi(\zeta)$ in previous figures. There remains a marked discrepancy for the most stable LES case with $h/L=$ 3.5, and local scaling underestimates the buoyancy effects for the conventionally neutral case with $h/L \ll$ 1. For the CASES-99 results, there is no discernible difference between $\Phi(z/\Lambda)$ and the previous results in Fig. \ref{fig3}(a,b). In both Figs. \ref{fig3} and \ref{fig4}, the proposed similarity parameter has lower NMSE values regardless of whether surface or local scaling is employed. The improved similarity with $Z$ in previous figures therefore cannot be explained by $Z$ better representing the local nondimensional parameters and their height-dependent properties.

As before, the mixed scaling parameter $z/\sqrt{\Lambda h}$ yields a complete collapse of the LES profiles in Fig. \ref{fig4}(c,d), suggesting mean profile similarity is better represented by $\Phi(z/\sqrt{\Lambda h})$ than $\Phi(z/\Lambda)$. The local scaling also appears to account for the weak curvature seen in Fig. \ref{fig2} and produces an unambiguously linear dependence. The collapse in Fig. \ref{fig4}(c,d) includes the traditional surface layer below $z=0.1h$ and the conventionally neutral case with top-down stability effects. In the geometric mean $\sqrt{L h}$, the depth $h$ has a strong modulating effect on extreme values of $L$: $h$ significantly decreases $Z$ for $h/L  \gg 1$ in strongly stable conditions (Fig. \ref{fig3}) and increases $Z$ for $h/L \ll 1$ in near-neutral conditions (Fig. \ref{fig4}). This modulation brings the results into alignment with the linear trend across the full range of $z/\sqrt{\Lambda h}$ tested here. In contrast, alternate combined length formulations such as an inverse summation length scale \citep[e.g.,][]{Delage1974,Zilitinkevich2005} always favor the small values and are less successful in capturing the trends observed here.

\subsection*{Dissipation}

To interpret the origin of the composite length scale $\sqrt{Lh}$, it is informative to consider the dissipation rate $\epsilon$ of TKE. For conditions in a local equilibrium between shear production $P= -\overline{u^\prime w^\prime} \, \partial \overline{U}/\partial z$, buoyancy production (destruction) $B = g \, \overline{w^\prime \theta^\prime}/\overline{\theta}$, and dissipation $\epsilon$, the nondimensional TKE budget reduces to \citep[e.g.,][]{Hartogensis2005}

\begin{equation}
0 = \Phi_m - \frac{z}{\Lambda} - \Phi_\epsilon,
\label{eq:tke}
\end{equation}

\noindent where $\Phi_m$ arises from production $P$, $z/\Lambda$ represents $B$, and $\Phi_\epsilon = \epsilon \kappa z / (-\overline{u^\prime w^\prime})^{3/2}$. Equation \eqref{eq:tke} can also be expressed in terms of surface scaling if the local fluxes in $P$ and $B$ are assumed to be equivalent to the surface parameters. Local fluxes are used here due to the observed flux decay throughout the surface layer. If $z/\sqrt{\Lambda h}$ is the appropriate similarity parameter for $\Phi_m$ as evidenced in Fig. \ref{fig4}, then Eq. \eqref{eq:tke} requires that $\Phi_\epsilon$ be a function of both $z/\sqrt{\Lambda h}$ and $z/\Lambda$.

\begin{figure*}
\centerline{\includegraphics[width=39pc]{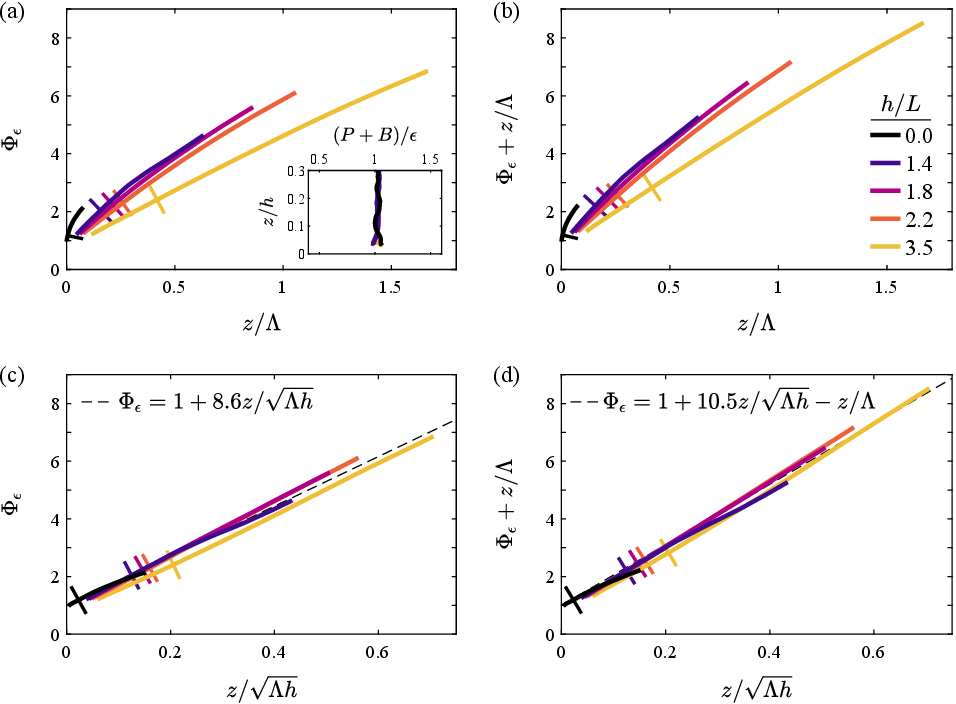}}
  \caption{Comparison of the dimensionless dissipation $\Phi_\epsilon= \epsilon \kappa z / (-\overline{u^\prime w^\prime})^{3/2}$ for the LES cases as a function of the traditional similarity parameter $z/\Lambda$ and the proposed parameter $z/\sqrt{\Lambda h}$, where $\Phi$ and $\Lambda$ are defined using local-in-height fluxes: (a) $\Phi_\epsilon(z/\Lambda)$; (b) $\Phi_\epsilon(z/\Lambda)+z/\Lambda$; (c) $\Phi_\epsilon(z/\sqrt{\Lambda h})$; and (d) $\Phi_\epsilon(z/\sqrt{\Lambda h})+z/\Lambda$. Vertical positions from $z=$ 0.03$h$ to 0.3$h$ are included here, and the short lines indicate the height $z=0.1h$ for reference. The results in (b) and (d) are offset by $z/\Lambda$ according to the turbulent kinetic energy (TKE) equilibrium condition in Eq. \eqref{eq:tke}. The inset panel of (a) shows the balance between TKE shear production $P$, buoyancy destruction $B$, and dissipation $\epsilon$. The relations in (c,d) result from a least-squares fit to the LES.}
  \label{fig5}
\end{figure*}

The dissipation dependence is evaluated in Fig. \ref{fig5} for the LES data. Dissipation was also estimate for the CASES-99 measurements using the amplitude of the energy spectrum in the inertial subrange \citep{Saddoughi1994}. While the local equilibrium is often approximately observed in the stable surface layer \citep{Wyngaard1971,Chamecki2018} due to increasingly large values of shear production and dissipation \citep{Wyngaard1992,Frenzen2001}, for many of the weakly stratified data periods an imbalance was observed between $P$, $B$, and $\epsilon$ such that $\Phi_\epsilon$ did not adhere to the expected trends. The imbalance may be due to a combination of nonstationarity in the TKE, contributions from transport terms that are neglected in Eq. \eqref{eq:tke}, and uncertainties in $\Phi_\epsilon$ estimates. Rather than selectively filtering the CASES-99 results to a smaller subset of the data that exhibits TKE equilibrium, the field measurements are excluded from Fig. \ref{fig5}.

The LES dissipation in Fig. \ref{fig5} is modeled as $\epsilon = C_\epsilon e^{3/2} / \Delta$, where $e$ is the subgrid-scale TKE and $C_\epsilon = 0.93$ \citep{Moeng1988,Sullivan2016}. The inset panel in Fig. \ref{fig5}(a) demonstrates the computed TKE equilibrium in the surface layer for each of the cases to support use of the simplified budget in Eq. \eqref{eq:tke}. The dissipation in Fig. \ref{fig5}(b,d) is offset by $z/\Lambda$ such that the resulting curves should match $\Phi_m$ from Fig. \ref{fig4}(a,c). The $h/L=3.5$ case in Fig. \ref{fig5}(c) is visibly displaced from the remaining curves, and the offset in Fig. \ref{fig5}(d) accounts for a majority of the observed displacement. Further, the result in Fig. \ref{fig5}(d) matches closely with Eq. \eqref{eq:tke} using the definition for $\Phi_m(z/\sqrt{\Lambda h})$ from the fitted relation in Fig. \ref{fig4}(c).

The observed similarity in Fig. \ref{fig5}(d) implies that the composite length scale $\sqrt{ \Lambda h}$ is specifically associated with the mean gradients $\partial \overline{U} / \partial z$ and $\partial \overline{\theta} / \partial z$. The scaling is also reflected by integral-scale turbulent features related to the mean gradients \citep{Heisel2023}, as will be discussed in Sec. \ref{sec:discuss}. The turbulent energy is predominately set by these integral-scale features and thus reflects the same mixed scaling as $\Phi_m$. However, a fraction of the integral-scale energy proportional to $z/\Lambda$ is directly dampened by negative buoyancy in stratified conditions. The remaining balance of the energy sets the required rate of small-scale dissipation in accordance with Eq. \eqref{eq:tke}, where the balance depends strongly on $\sqrt{\Lambda h}$ through the shear production and has a weaker dependence on $\Lambda$ owing to direct buoyancy effects. In other words, the rate of small-scale dissipation reflects a mixture of nondimensional similarity parameters resulting from competing large-scale effects.

\section{Discussion}
\label{sec:discuss}

\subsection*{The limit of z-less stratification}

The results in Sec. \ref{sec:results} indicate the relevance of $h$ to mean profile similarity in the surface layer, but the evidence thus far does not indicate \textit{why} this should be the case. In fact, the idea that the surface layer is far from the influence of the upper boundary (i.e. $z \ll h$) is a common argument for excluding $h$ as a relevant length scale in derivations of the log law. It is speculated here that this dependence originates from the limit of the mean gradients in z-less stratification.

Assuming a linear similarity relation for the SBL, the velocity gradient can be expressed as

\begin{equation}
\frac{\partial \overline{U}}{\partial z} \approx \frac{u_*}{\kappa z} +\left( \frac{\partial \overline{U}}{\partial z} \right) _{z-less}.
\label{eq:grad}
\end{equation}

\begin{figure*}
\centerline{\includegraphics[width=39pc]{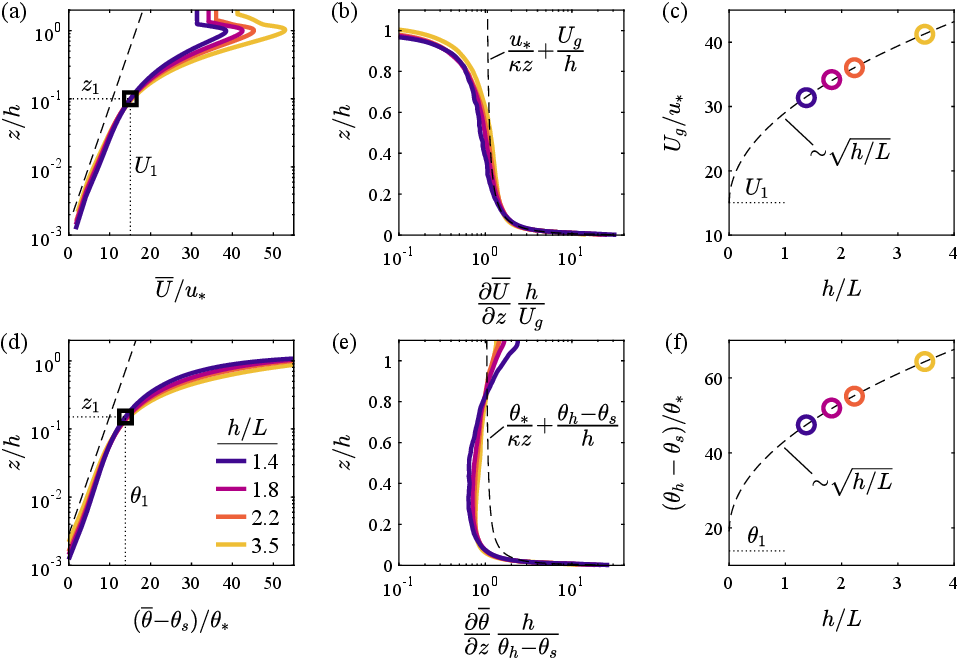}}
  \caption{Estimation of the bulk z-less gradient above the near-surface region: (a,d) approximate position where the mean profiles strongly deviate from log-linear scaling; (b,e) mean gradients relative to the bulk z-less approximation; (c,f) bulk difference in velocity and temperature as a function of stability $h/L$. The rows correspond to velocity (a,b,c) and temperature (d,e,f), where $U_g$ is the geostrophic wind speed, $\theta_h$ is the temperature at $z=h$, $\theta_s$ is the surface temperature, and subscript ``1'' indicates the properties at the estimated deviation points in the profiles. The dashed lines represent $\log(z)$ in (a,d), Eqs. \eqref{eq:drag} and \eqref{eq:dragtemp} in (b,e) and $\sqrt{h/L}$ in (c,f).}
  \label{fig6}
\end{figure*}

\noindent Log law scaling is approximately achieved near the surface for small $z$ when $u_*/\kappa z$ is the dominant term. Far from the surface and under strongly stratified conditions, the log law gradient becomes relatively small and the result asymptotically approaches the z-less gradient with increasing height. This scaling behavior is also manifested by coherent features in the instantaneous turbulence. The size of the coherent regions is proportional to $z$ in near-neutral conditions \citep{Heisel2018,Heisel2020} and decreases towards a constant value as stratification increases \citep{Heisel2023}. Equation \eqref{eq:grad} is further supported by the LES profiles featured in Fig. \ref{fig6}. Both the mean profiles in Fig. \ref{fig6}(a,d) and their gradients in Fig. \ref{fig6}(b,e) exhibit a strong dependence on $z$ in the surface layer, and the gradients are approximately z-less throughout a large portion of the outer layer.

For MOST, the z-less gradient in Eq. \eqref{eq:grad} is proportional to $u_*/L$. While $L$ quantifies the height at which buoyancy effects are important, it does not necessitate that the turbulence and statistics be directly proportional to $L$ at any height. In this context, alternative scales such as the Ozmidov length have been proposed for stable conditions \citep{Li2016}. A simpler prediction is that the z-less gradient is constant throughout the depth of the SBL, leading to the value

\begin{equation}
\left( \frac{\partial \overline{U}}{\partial z} \right) _{z-less} \approx \frac{U_g}{h},
\label{eq:zless}
\end{equation}

\noindent where $U_g$ is the geostrophic wind speed that describes the bulk velocity difference across $h$. While the speed of the low-level jet (LLJ) at height $h$ is expected to be more representative than $U_g$ in Eq. \eqref{eq:zless}, the geostrophic wind speed is used here in order to relate the results to existing resistance laws. Substituting the Eq. \eqref{eq:zless} approximation into Eq. \eqref{eq:grad} yields

\begin{equation}
\frac{\partial \overline{U}}{\partial z} \approx \frac{u_*}{\kappa z} \left( 1 + \kappa \frac{U_g}{u_*} \frac{z}{h} \right),
\label{eq:drag}
\end{equation}

\noindent where $u_*/U_g$ is the geostrophic drag coefficient predicted from resistance laws \citep{Rossby1935,Lettau1959}. The same z-less approximation can be made for the temperature, resulting in

\begin{equation}
\frac{\partial \overline{\theta}}{\partial z} \approx \frac{\theta_*}{\kappa z} \left( 1 + \kappa \frac{\theta_h - \theta_s}{\theta_*} \frac{z}{h} \right)
\label{eq:dragtemp}
\end{equation}

\noindent based on the temperatures $\theta_h$ at $z=h$ and $\theta_s$ at the surface.

Equations (\ref{eq:drag}, \ref{eq:dragtemp}) are represented by dashed lines in Fig. \ref{fig6}(b,e). The equations align well with the general shape of the profiles. The coarse z-less approximation employed here overpredicts the temperature gradient by 20-30\% in Fig. \ref{fig6}(e), but this difference can be accounted for using a constant correction factor. Based on Fig. \ref{fig6}(b,e), the relevance of $h$ in the similarity relations results from an approximately z-less gradient directly above the surface layer that creates an asymptotic boundary condition expressed through Eq. \eqref{eq:grad}.

The connection between the approximations in Eqs. \eqref{eq:drag} and \eqref{eq:dragtemp} and the composite length scale $\sqrt{Lh}$ relies on the resistance laws for drag $U_g/u_*$ and heat transfer $(\theta_h-\theta_s)/\theta_*$. The stability dependence of these laws is evaluated in Fig. \ref{fig6}(c,f). The trends closely match $\sqrt{h/L}$ for both velocity and temperature, such that the z-less terms in Eqs. \eqref{eq:drag} and \eqref{eq:dragtemp} can be approximated as

\begin{equation}
\begin{split}
\kappa \frac{U_g}{u_*} \frac{z}{h} \approx \kappa C_u \frac{z}{\sqrt{Lh}} \\
\kappa \frac{\theta_h-\theta_s}{\theta_*} \frac{z}{h} \approx \kappa C_\theta \frac{z}{\sqrt{Lh}}.
\end{split}
\label{eq:quad}
\end{equation}

\noindent The dependence of the approximate z-less gradient on stratification therefore results in the same revised similarity parameter as Eq. \eqref{eq:heist}. Based on the fitted coefficients $C_u=$ 14 and $C_\theta=$ 24 in the Fig. \ref{fig6}(c,f) curves, the constant factors in Eq. \eqref{eq:quad} are 5.6 and 9.7 for velocity and temperature, respectively. These values are reasonably similar to the slope factors in Eq. \eqref{eq:heist} in consideration of the coarse assumptions made here.

One notable assumption for the bulk z-less gradient is to neglect the effect of $z$ scaling in the surface layer where the gradients are significantly greater than the z-less approximation. Excluding the surface layer from the z-less gradient leads to the revised approximations $(U_g-U_1)/(h-z_1)$ and $(\theta_h-\theta_1)/(h-z_1)$, where $U_1$ and $\theta_1$ correspond to the estimated departure points from $z$ scaling shown in Fig. \ref{fig6}(a,d). These surface layer offsets $U_1$ and $\theta_1$ account for a majority of the fitted intercepts in Fig. \ref{fig6}(c,f), respectively, and help to explain the overprediction and small discrepancies seen in the outer layer in Fig. \ref{fig6}(b,e).

Numerous previous works have studied the effect of stable stratification on $U_g/u_*$ and $(\theta_h-\theta_s)/\theta_*$, where the stability parameter is typically $\mu = u_*/fL$ or $\mu = h/L$ \citep{Melgarejo1974}. Many of the derived dependencies include a combination of $\log{\mu}$ and either $\mu$ or $\sqrt{\mu}$ \citep[e.g.,][]{Yamada1976,Arya1977,Duvachat1982,Byun1991,Zilitinkevich2005}. The existence of resistance law functions with $\sqrt{\mu}$ provides a precedent for the present observations. While the current results suggest that $\sqrt{h/L}$ provides a good approximation for the resistance, the results are not conclusive due to the limited number of points in Fig. \ref{fig6}(c,f) and the complex parameter space of the resistance laws. In this context, the coefficients in Eqs. \eqref{eq:heist} and \eqref{eq:quad} may not be universal. Specifically, the fitted slopes may depend on additional factors such as $f$ and $N$ that are fixed across the present LES cases.

\begin{figure}
\centerline{\includegraphics[width=19pc]{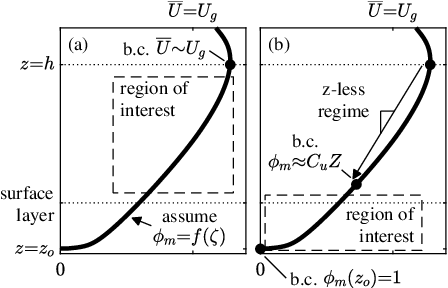}}
  \caption{Comparison of the traditional (a) and proposed (b) approach for matching similarity profiles in the ABL. In (a), profiles in the outer layer are designed to match MOST in the surface layer and satisfy the upper boundary condition (b.c.) at $z=h$ \citep[e.g.,][]{Byun1991}. In (b), the gradient in Eq. \eqref{eq:quad} is observed in the vicinity of the surface region due to approximately z-less stratification in the outer layer, such that the effect of $h$ (through $Z$) cannot be neglected in the surface layer. The profile shown is the LES case with $h/L =$ 3.5.}
  \label{fig7}
\end{figure}

Another important consideration is that many derivations of the resistance laws invoke MOST in the surface layer as a boundary condition in order to determine the mean profiles aloft, such that a direct quantitative comparison cannot be made between the previous literature and the present discussion. This difference in approach is visualized in Fig. \ref{fig7}. Traditional derivations depicted in Fig. \ref{fig7}(a) use MOST relations in the surface layer and an upper boundary condition at $z=h$ to devise defect forms of the velocity and temperature profiles within the outer layer of the ABL \citep[see, e.g.,][]{Byun1991}.

For the current approach shown in Fig. \ref{fig7}(b), the presence of z-less stratification and an approximately constant gradient throughout a majority of the outer layer lead to a known matching condition in the vicinity of the surface layer. The boundary condition is no longer far from the surface layer, such that the profiles in the surface layer must consider the influence of $\sqrt{Lh}$ associated with the z-less gradient. The gradient profile in Eq. \eqref{eq:grad} matches both the surface condition $\phi_m \approx$ 1 and the z-less gradient in the respective asymptotic limits $Z \ll 1$ and $Z \gg 1$, where the linear form of the similarity relation is supported by the evidence shown here and in the literature.

The proposed explanation for mixed scaling outlined in Fig. \ref{fig7}(b) applies also to the conventionally neutral LES case. A large portion of the outer layer in this case exhibits an approximately z-less mean velocity gradient owing to the top-down buoyancy effects \citep[see, e.g., Fig. 1 of][]{Heisel2023}, where the flux Richardson number exceeds 0.1 throughout the top half of the ABL. The conventionally neutral case is thus expected to have the same z-less matching condition as the stable LES, except local scaling is required to account for the effect of the top-down stratification as seen in Fig. \ref{fig4}.

\subsection*{The shape of $\phi(\zeta)$}

We return now to Fig. \ref{fig1} to consider the non-linear shape of $\phi(\zeta)$ and variability between fitted relations in the context of the proposed mixed length similarity. Assuming $Z$ and the linear fit in Eq. \eqref{eq:heist} are the correct choice for mean profile similarity in the SBL, the dependence of the proposed similarity relations on $\zeta$ is

\begin{equation}
\phi \left(\zeta,\frac{h}{L} \right)  = A + B \zeta \left( \frac{h}{L} \right) ^{-1/2}
\label{eq:project}
\end{equation}

\noindent for both momentum and heat, where $A$ and $B$ are assumed to be constants. The result is a linear relation for a given stability $h/L$ whose slope decreases with increasing stability as seen in Fig. \ref{fig2}(a,b).

\begin{figure*}
\centerline{\includegraphics[width=39pc]{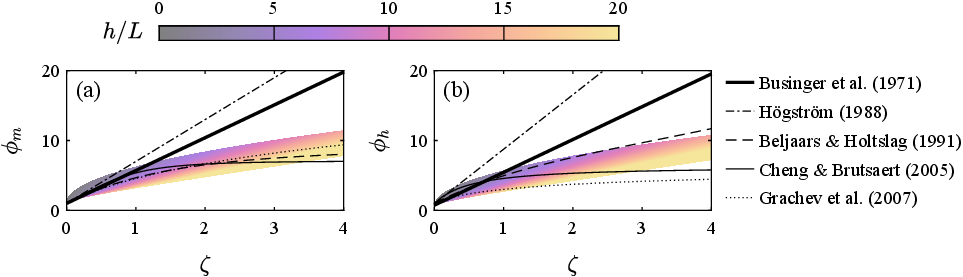}}
  \caption{Same as Fig. \ref{fig1}, with the addition of $\phi(Z)$ curves projected onto the $\zeta{-}\phi$ parameter space for a range of $h/L$ conditions (color) using the relations in Eq. \eqref{eq:heist}. The resulting $\phi(\zeta)$ curves for a fixed $h/L$ value are linear, but the linear trend is masked by the limited range of $z/h$ that can be observed from typical field measurements (between $z=$ 1.5 m and 0.3$h$ in this example). The non-linear shape of the colored band -- resulting from the connection between $Z$, $\zeta$, $h$, and $L$ -- closely resembles typical $\phi(\zeta)$ curves for strongly stable conditions.}
  \label{fig8}
\end{figure*}

In practice, the range of heights observed for a given stability $h/L$ is selectively sampled based on measurement limitations. The lower bound of the range is given by either the lowest anemometer height or the depth of the roughness sublayer, and the upper bound is determined by either the tower height or the surface layer depth for shallow boundary layers. The effect of this range is visualized in Fig. \ref{fig8} which shows Eq. \eqref{eq:project} with $h/L$ indicated by color and the solution confined to heights between $z=$ 1.5 m and 0.3$h$. Here, the SBL depth in meters is assumed to be $h = 1.25 (40 - h/L)$ based on a weak linear trend observed from the CASES-99 results for large $h/L$, where the leading constant 1.25 has units of meters.

The selective height sampling discussed above acts as a bandpass filter on the solution space in Fig. \ref{fig8}, where the shape of the resulting band resembles the non-linear empirical relations for $\phi(\zeta)$. This shape is determined by multiple factors. The most notable is the square root relation in Eq. \eqref{eq:project}, leading to $\phi \sim \sqrt{\zeta}$ for a fixed $z/h$. The shape also depends on the relation between $h$ and $L$ for a given field site and the specific heights of the field instruments. The limits of the band move to lower values of $\phi$ if the measurement point is closer to the ground relative to $h$, which can correspond to either a lower anemometer position or a deeper SBL for a given $L$. In the same way, a higher measurement point or shallower SBL leads to higher observed $\phi$ values. The projection of the proposed similarity in Fig. \ref{fig8} and its multiple dependencies may help to explain the observed non-linearity in $\phi(\zeta)$ and some of the variability across different field campaigns.

\section{Conclusion}
\label{sec:conclude}

The evidence presented here supports a composite length scale $\sqrt{Lh}$ for similarity in the mean profiles of wind speed and air temperature under stable stratification. The mixed scaling combines the additional dependence on $h$ into a single parameter $Z$ that achieves improved similarity in mean gradients from LES of the SBL (Fig. \ref{fig2}). This similarity extends to LES of a conventionally neutral ABL when local-in-height scaling is considered (Fig. \ref{fig4}). For dissipation $\epsilon$ under equilibrium conditions, the similarity relations depend strongly on $Z$ and weakly on $\zeta$ (Fig. \ref{fig5}) in accordance with the TKE budget (Eq. \ref{eq:tke}). While traditional MOST and $\zeta$ can be used to predict a majority of the deviation from log law scaling in the mean gradients and dissipation, the proposed similarity based on $Z$ accounts for the remaining differences and matches closely with the profile trends as seen in Table \ref{tbl2}.

The linear similarity relations in Eq. \eqref{eq:heist} fitted to the LES profiles align well with field measurements from the CASES-99 experiment, including for strong stability where a departure from MOST is often observed (Fig. \ref{fig3}). The shape of this departure in $\phi(\zeta)$ can be explained by the connection between $Z$ and $\zeta$ (Eq. \ref{eq:project}) and the limited range of heights that can be sampled in experiments (Fig. \ref{fig8}).

It is speculated that the relevance of $h$ to mean profile similarity in the surface layer may be related to approximately z-less stratification above the surface layer (Fig. \ref{fig6}), where the z-less limit may impose a boundary condition that influences the surface layer flow structure (Fig. \ref{fig7}). This dependence can be represented functionally by blending the log law with a z-less limit defined using upper boundary parameters (Eqs. \ref{eq:drag} and \ref{eq:dragtemp}). The proposed mixed scaling emerges from these functions after applying observed resistance relations (Eq. \ref{eq:quad}).

The conceptual explanation of the mixed scaling and the similarity relations in Eq. \eqref{eq:heist} are considered to be preliminary efforts towards accounting for the trends observed here. Both the LES profiles and CASES-99 measurements support the linear relations

\begin{equation}
\begin{split}
\phi_m(Z) &= a_m + b_m Z \\
\phi_h(Z) &= a_h + b_h Z \\
\phi_\epsilon(Z,\zeta) &= a_m + b_m Z - \zeta, \\
\end{split}
\label{eq:final}
\end{equation}

\noindent where $a_m=1$ is imposed to match the log law scaling for neutral conditions and $\phi_\epsilon$ is strictly valid under local equilibrium of TKE. Yet, there is uncertainty in the value of the fitted parameters $b_{m,h}$ and $a_h$. The optimal values for the LES profiles are $b_m \approx 9.5{-}10.5$, $b_h \approx 8.4{-}9.3$, and $a_h\approx0.55{-}0.72$, which all depend on the range of heights included in the fit and whether surface or local scaling is employed. If the arguments in Sec. \ref{sec:discuss} are valid, the slopes $b_{m,h}$ may exhibit some dependence on additional parameters relevant to the resistance laws such as $f$ and $N$ that were not tested here.

Further, the LES and field experiments both represent idealized stationary conditions and surface-forced turbulence. For instance, the CASES-99 data periods were selected based on idealized flux profiles that result from assumptions of stationarity and a constant Richardson number profile \citep{Nieuwstadt1984}. These conditions are more common to long-lived boundary layers in polar and marine environments than traditional nocturnal boundary layers. The generality of the mixed scaling should therefore be corroborated and refined in future analyses, with specific scrutiny given to testing transient flow conditions, the appropriate intercept $a_h$, and the universality of $b_{m,h}$. However, it is expected that the evaluation of additional field measurements will be constrained by the requirement for a reliable estimation of the SBL depth.

Direct knowledge of $h$ in the definition of $Z$ can be avoided by using derived relations that consider the effects of rotation, stratification above the boundary layer, and surface-forced buoyancy \citep{Zilitinkevich1996,Mironov2010}. The limiting cases include $h \sim u_*/f$ for a neutral Ekman layer \citep{Rossby1935}, $h \sim u_*/N$ for a conventionally neutral boundary layer with an overlying capping inversion \citep{Kitaigorodskii1988}, and $h \sim L$ for pure surface forcing in the absence of rotation \citep{Kitaigorodskii1960}. While the last relation is derived invoking MOST in the surface layer, the same dependence occurs if $L$ is replaced by $\sqrt{hL}$, suggesting that the mixed scaling may reduce to traditional Monin-Obukhov similarity under specific simplified conditions. Importantly, each of these estimates represents the equilibrium depth. Under nonstationary conditions, the true depth -- and the applicability of the mixed scaling as noted above -- become more complicated \citep{Nieuwstadt1981}.

Finally, it is worth noting that the revised similarity can alternatively be expressed through multiple dimensionless parameters, e.g. $\phi = f(\zeta,h/L)$ as seen in Eq. \eqref{eq:project}. The presence of multiple dependencies is consistent with the transition between stability regimes discussed in the introduction, but does not comply with Buckingham $\pi$ theorem \citep{Buckingham1914}. Rather, it is a case of \textit{similarity of the second kind} \citep{Barenblatt2003}, where the exponent for each parameter cannot be determined directly by dimensional analysis. Here, the exponents in Eq. \eqref{eq:project} were determined through empirical observation and are additionally supported by the physical arguments and results presented in Sec. \ref{sec:discuss}. These exponents are simplified into the single parameter $Z$ in order to reduce the dimensionality of the similarity relations.

\acknowledgments
M. H. gratefully acknowledges funding support from the US National Science Foundation (NSF-AGS-2031312). M. C. is supported by the Biological and Environmental Research program of the Department of Energy (DE-SC0022072). The authors are grateful to several colleagues for discussions and suggestions related to the present work: N. L. Dias regarding incomplete similarity, G. G. Katul regarding the gradients in the limit of z-less stratification, and P. P. Sullivan regarding top-down buoyancy effects. The authors are additionally thankful to P. P. Sullivan for sharing the LES results.

\datastatement
The unprocessed CASES-99 measurements can be accessed from a public database maintained by the Earth Observing Laboratory of the National Center for Atmospheric Research. The database includes met tower measurements from the sonic anemometers (\url{https://doi.org/10.5065/D6H993GM}), vane anemometers and thermistors (\url{https://doi.org/10.5065/D67W69HS}), and thermocouples (\url{https://doi.org/10.26023/V9XE-59MN-AH01}) used in the present analysis. The LES profiles for mean wind speed and air temperature are included in a repository for a previous study (\url{https://doi.org/10.26023/V9XE-59MN-AH01}).

\appendix

\appendixtitle{Evaluation of CASES-99 field measurements}

The CASES-99 database includes continuous met tower measurements spanning the duration of October 1999. The measurements were segmented into 5-minute intervals to be individually evaluated and processed. Only nighttime data periods between 00:00 and 12:00 coordinated universal time (7 pm to 7 am local time) were considered due to the higher likelihood of stably-stratified conditions during these hours. The following additional criteria were evaluated for each 5-minute period:

\begin{enumerate}

\item The mean wind direction during the period was between 90 to 270 degrees, i.e. the wind was from east to west. The heading of the anomemeters towards the east requires a westerly wind for reliable measurements. The relatively wide $\pm$90$^\circ$ allowable wind range was chosen in consideration of the predominant wind directions of approximately 90 and 270 degrees observed across all measurements.

\item No more than 5\% of the data points were identified as outliers. Outliers include points flagged directly by the anemometer and points identified from the spike detection and removal filter detailed in \citet{Vickers1997}. All outliers were replaced by linear interpolation.

\item No more than 25\% change in mean wind speed and direction across the 5-minute period. The nonstationarity in mean conditions was determined from linear regression of the time series \citep{Vickers1997}.

\end{enumerate}

The data period was accepted if at least six of the eight sonic anemometers met all three criteria listed above. A majority of anemometers are required to ensure the flux profile fit detailed below produces a reliable result. For computing fluxes, turbulent fluctuations were estimated by subtracting the linear regression of each sonic anemometer time series.

The parameters $h$, $u_*$, and $\theta_*$ were then estimated from a nonlinear least-squares fit of $\overline{u^\prime w^\prime}$ and $\overline{w^\prime \theta^\prime}$ to idealized flux profiles. The idealized profiles are assumed to be piecewise functions with $-\overline{u^\prime w^\prime} = u_*^2 (1-z/h)^{3/2}$ and $- \overline{w^\prime \theta^\prime} = u_* \theta_* (1-z/h)$ for $z \le h$ \citep{Nieuwstadt1984}, and $\overline{u^\prime w^\prime} = \overline{w^\prime \theta^\prime} = 0$ for $z>h$. The cost function for the fit is the residual between the computed fluxes and the piecewise profiles, where the residual considers both the momentum and kinematic heat fluxes simultaneously. Combining the two fluxes in the total residual ensures that the outputted parameters reflect both measured flux profiles.

Another consideration for the fitting procedure is the initial guesses for $h$, $u_*$, and $\theta_*$. The output was found to be somewhat sensitive to arbitrary initial values, such that informed guesses are required. The initial values for $u_*$ and $\theta_*$ are taken from the measured fluxes at the lowest height, and $h$ is taken to be either the first height where the fluxes are less than 25\% of the surface value or the top of the tower, whichever is lower. 

\begin{figure}
\centerline{\includegraphics[width=19pc]{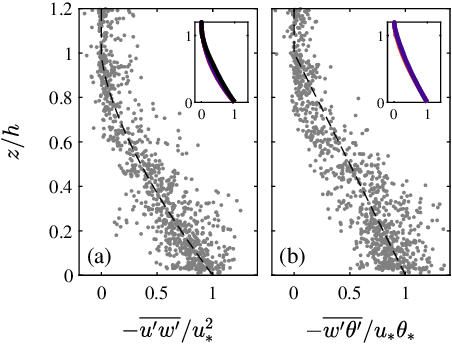}}
  \caption{Composite of dimensionless flux profiles for all accepted 5-minute data periods from the CASES-99 field measurements: (a) momentum and (b) kinematic heat. The parameters $h$, $u_*$, and $\theta_*$ for each data period result from a least-squares fit to the profiles $-\overline{u^\prime w^\prime}/u_*^2 = (1-z/h)^{3/2}$ and $-\overline{w^\prime \theta^\prime}/u_* \theta_* = (1-z/h)$ (dashed lines), where zero flux is assumed above $h$. The inset panels show the corresponding LES profiles (solid lines).}
  \label{figA1}
\end{figure}

Figure \ref{figA1} shows the nondimensional flux profiles resulting from the fitting procedure, where the dashed lines represent the idealized piecewise functions. The choice of the fitted functions is supported by the LES flux profiles shown in the inset panels, where the profiles overlap closely with the piecewise functions such that the dashed lines are not visible. Revising the piecewise functions to assume a small non-zero flux above $h$ (e.g. 5\% of the surface value) does not change the conclusions of the analysis.

The fitted results were only accepted if $R^2>0.7$ based on the coefficient of determination. Importantly, the fitting procedure and $R^2$ threshold favor shallow boundary layers for which there is substantial flux decay across the height of the tower. For deeper boundary layers, it becomes harder to distinguish the more gradual flux decay from random error in the fluxes, leading to lower $R^2$ values that do not pass the threshold. This limitation can be accepted for the present analysis because the trends of interest occur in the strongly stable regime that is typically associated with a shallower SBL.

In addition to the $R^2$ criterion, minimum values were imposed for $h$, $u_*$, and $\theta_*$. Fits with $h <$ 10 m were excluded to avoid periods with a collapsed boundary layer and to ensure at least three sonic anemometers were positioned within the SBL. Small flux values $u_*^2 <$ 0.0004 m$^2$\,s$^{-2}$ and $u_* \theta_* <$ 0.0004 K\,m\,s$^{-1}$ were also excluded to avoid periods with very weak or intermittent turbulence. The sonic anemometers were a mixture of CSAT3 and ATI-K sonics, with the latter positioned at 10, 20, 40, and 55 m. The ATI-K sonics were found to have digitized values with a resolution of 0.01 for both velocity (m\,s$^{-1}$) and temperature (K). The CSAT3 sonics were rounded to the same resolution to compare computed fluxes from the coarsened signal to the original result. The relative error in the fluxes due to the digital resolution was less than 10\% when the fluxes exceeded the minimum values above, which guided the selection of the threshold.

\begin{figure}
\centerline{\includegraphics[width=19pc]{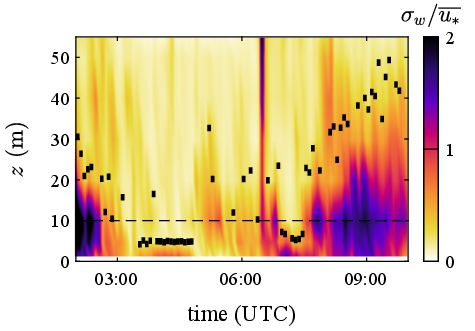}}
  \caption{Example space-time contour of the vertical root-mean-square velocity $\sigma_w$ from the CASES-99 field measurements during a ten hour period on 18 October 1999, overlaid with the estimated boundary layer depth $h$ for individual 5-minute data periods. The values are shown relative to the average shear velocity $\overline{u_*}$ for the time period shown. The dashed line at $z=$ 10 m is the minimum $h$ considered in this study.}
  \label{figA2}
\end{figure}

An example space-time contour of turbulent statistics for a ten-hour period is shown in Fig. \ref{figA2}, where the statistics are computed in 5-minute averaging intervals. The figure demonstrates how the fitted SBL depth $h$ aligns well with the vertical position where the turbulent fluctuations become small relative to $u_*$. Figure \ref{figA2} also illustrates multiple periods when turbulence collapses above the first measurement point such that the fitted depth is below the minimum threshold, i.e. the dashed line.

To estimate the gradient statistics, the mean profiles were fitted to the second-order polynomial $A + B \log{(z)} + C \log{(z)}^2$ and the derivatives were computed as $B/z + 2 C \log{(z)} / z$ \citep{Hogstrom1988}. The mean velocity profiles used both the sonic and vane anemometers due to an observed agreement between the two instruments. The thermocouples were used for the mean temperature profiles to avoid drift issues associated with the sonic anemometers. The thermistor measurements were also used during occasional periods when the thermocouples were unavailable. The least-squares polynomial fits only considered points in the lowest 30\% of the SBL relative to the fitted $h$. However, a minimum of four points was used in the case of a shallow SBL to avoid applying a 3-parameter fit to three or fewer points. Mean profile fits with $R^2 < 0.9$ are excluded from the results presented here. From the fitted polynomial, the gradients were evaluated at the heights of the sonic anemometers to support the comparison of local-in-height scaling shown in Fig. \ref{fig4}.

Energy spectra were inspected for several of the accepted 5-minute data subsets under strongly stable conditions. The spectra exhibited a canonical inertial subrange power law despite many of the periods having an estimated flux Richardson number above the critical value $\mathrm{Ri}_f \approx 0.2$ \citep{Grachev2013}. The discrepancy may be due to uncertainty in the estimated $\mathrm{Ri}_f$ values or the presence of non-local turbulent energy sources such as transport \citep{Freire2019}.

Out of 7200 5-minute data subsets evaluated, 205 (2.9\%) subsets met all the criteria above. Several tests were conducted to ensure the trends in Fig. \ref{fig3} are not sensitive to the methodology. These tests include longer data periods, changes to the threshold values for maximum nonstationarity and profile fit $R^2$, finite difference estimates for gradients along $\log{(z)}$, and different methods for computing fluctuations. The use of longer data periods and stricter nonstationarity tests such as the reverse arrangement test \citep{Dias2004}, in combination with the flux profile fits, were found to eliminate too many data periods to discern any meaningful trend. Determination of a spectral gap and application of a high-pass filter to estimate fluxes \citep{Vickers2003} modestly reduces the scatter in the weakly stable regime, but the method employed here is preferred due to its simplicity. While the various tests changed the number of accepted data points and to a lesser extent the scatter, none of the tests yielded trends opposing the findings in Sec. \ref{sec:results}.

Finally, the depth $h$ was alternatively estimated using lidar data and the height of detected low-level jets (LLJ) \citep{Banta2002}. However, the LLJ was often far above the decay of fluxes and plateau of the mean profiles along the main met tower, consistent with previous findings \citep{Banta2007}. The observation suggests that the lowest LLJ that can be detected by the lidar is not always representative of the SBL depth, especially for strong stratification with a shallow SBL, such that the tower profile fits in Fig. \ref{figA1} are preferred for the present study.

\bibliographystyle{ametsocV6}
\bibliography{references}

\end{document}